\documentstyle{l-aa}
\def\PsfigVersion{1.9}
\ifx\undefined\psfig\else \fi

\let\LaTeXAtSign=\@
\let\@=\relax
\edef\psfigRestoreAt{\catcode`\@=\number\catcode`@\relax}
\catcode`\@=11\relax
\newwrite\@unused
\def\ps@typeout#1{{\let\protect\string\immediate\write\@unused{#1}}}
\ps@typeout{psfig/tex \PsfigVersion}

\def\figurepath{./}

\def\@nnil{\@nil}
\def\@empty{}
\def\@psdonoop#1\@@#2#3{}
\def\@psdo#1:=#2\do#3{\edef\@psdotmp{#2}\ifx\@psdotmp\@empty \else
    \expandafter\@psdoloop#2,\@nil,\@nil\@@#1{#3}\fi}
\def\@psdoloop#1,#2,#3\@@#4#5{\def#4{#1}\ifx #4\@nnil \else
       #5\def#4{#2}\ifx #4\@nnil \else#5\@ipsdoloop #3\@@#4{#5}\fi\fi}
\def\@ipsdoloop#1,#2\@@#3#4{\def#3{#1}\ifx #3\@nnil 
       \let\@nextwhile=\@psdonoop \else
      #4\relax\let\@nextwhile=\@ipsdoloop\fi\@nextwhile#2\@@#3{#4}}
\def\@tpsdo#1:=#2\do#3{\xdef\@psdotmp{#2}\ifx\@psdotmp\@empty \else
    \@tpsdoloop#2\@nil\@nil\@@#1{#3}\fi}
\def\@tpsdoloop#1#2\@@#3#4{\def#3{#1}\ifx #3\@nnil 
       \let\@nextwhile=\@psdonoop \else
      #4\relax\let\@nextwhile=\@tpsdoloop\fi\@nextwhile#2\@@#3{#4}}
\ifx\undefined\fbox
\newdimen\fboxrule
\newdimen\fboxsep
\newdimen\ps@tempdima
\newbox\ps@tempboxa
\long\def\fbox#1{\leavevmode\setbox\ps@tempboxa\hbox{#1}\ps@tempdima\fboxrule
    \advance\ps@tempdima \fboxsep \advance\ps@tempdima \dp\ps@tempboxa
   \hbox{\lower \ps@tempdima\hbox
  {\vbox{\hrule height \fboxrule
          \hbox{\vrule width \fboxrule \hskip\fboxsep
          \vbox{\vskip\fboxsep \box\ps@tempboxa\vskip\fboxsep}\hskip 
                 \fboxsep\vrule width \fboxrule}
                 \hrule height \fboxrule}}}}
\fi
\newread\ps@stream
\newif\ifnot@eof       
\newif\if@noisy        
\newif\if@atend        
\newif\if@psfile       
{\catcode`\%=12\global\gdef\epsf@start{
\def\epsf@PS{PS}
\def\epsf@getbb#1{%
\openin\ps@stream=#1
\ifeof\ps@stream\ps@typeout{Error, File #1 not found}\else
   {\not@eoftrue \chardef\other=12
    \def\do##1{\catcode`##1=\other}\dospecials \catcode`\ =10
    \loop
       \if@psfile
	  \read\ps@stream to \epsf@fileline
       \else{
	  \obeyspaces
          \read\ps@stream to \epsf@tmp\global\let\epsf@fileline\epsf@tmp}
       \fi
       \ifeof\ps@stream\not@eoffalse\else
       \if@psfile\else
       \expandafter\epsf@test\epsf@fileline:. \\%
       \fi
          \expandafter\epsf@aux\epsf@fileline:. \\%
       \fi
   \ifnot@eof\repeat
   }\closein\ps@stream\fi}%
\long\def\epsf@test#1#2#3:#4\\{\def\epsf@testit{#1#2}
			\ifx\epsf@testit\epsf@start\else
\ps@typeout{Warning! File does not start with `\epsf@start'.  It may not be a PostScript file.}
			\fi
			\@psfiletrue} 
{\catcode`\%=12\global\let\epsf@percent=
\long\def\epsf@aux#1#2:#3\\{\ifx#1\epsf@percent
   \def\epsf@testit{#2}\ifx\epsf@testit\epsf@bblit
	\@atendfalse
        \epsf@atend #3 . \\%
	\if@atend	
	   \if@verbose{
		\ps@typeout{psfig: found `(atend)'; continuing search}
	   }\fi
        \else
        \epsf@grab #3 . . . \\%
        \not@eoffalse
        \global\no@bbfalse
        \fi
   \fi\fi}%
\def\epsf@grab #1 #2 #3 #4 #5\\{%
   \global\def\epsf@llx{#1}\ifx\epsf@llx\empty
      \epsf@grab #2 #3 #4 #5 .\\\else
   \global\def\epsf@lly{#2}%
   \global\def\epsf@urx{#3}\global\def\epsf@ury{#4}\fi}%
\def\epsf@atendlit{(atend)} 
\def\epsf@atend #1 #2 #3\\{%
   \def\epsf@tmp{#1}\ifx\epsf@tmp\empty
      \epsf@atend #2 #3 .\\\else
   \ifx\epsf@tmp\epsf@atendlit\@atendtrue\fi\fi}

\chardef\psletter = 11 
\chardef\other = 12

\newif \ifdebug 
\newif\ifc@mpute 
\c@mputetrue 

\let\then = \relax
\def\r@dian{pt }
\let\r@dians = \r@dian
\let\dimensionless@nit = \r@dian
\let\dimensionless@nits = \dimensionless@nit
\def\internal@nit{sp }
\let\internal@nits = \internal@nit
\newif\ifstillc@nverging
\def \Mess@ge #1{\ifdebug \then \message {#1} \fi}

{ 
	\catcode `\@ = \psletter
	\gdef \nodimen {\expandafter \n@dimen \the \dimen}
	\gdef \term #1 #2 #3%
	       {\edef \t@ {\the #1}
		\edef \t@@ {\expandafter \n@dimen \the #2\r@dian}%
		\t@rm {\t@} {\t@@} {#3}%
	       }
	\gdef \t@rm #1 #2 #3%
	       {{%
		\count 0 = 0
		\dimen 0 = 1 \dimensionless@nit
		\dimen 2 = #2\relax
		\Mess@ge {Calculating term #1 of \nodimen 2}%
		\loop
		\ifnum	\count 0 < #1
		\then	\advance \count 0 by 1
			\Mess@ge {Iteration \the \count 0 \space}%
			\Multiply \dimen 0 by {\dimen 2}%
			\Mess@ge {After multiplication, term = \nodimen 0}%
			\Divide \dimen 0 by {\count 0}%
			\Mess@ge {After division, term = \nodimen 0}%
		\repeat
		\Mess@ge {Final value for term #1 of 
				\nodimen 2 \space is \nodimen 0}%
		\xdef \Term {#3 = \nodimen 0 \r@dians}%
		\aftergroup \Term
	       }}
	\catcode `\p = \other
	\catcode `\t = \other
	\gdef \n@dimen #1pt{#1} 
}

\def \Divide #1by #2{\divide #1 by #2} 

\def \Multiply #1by #2
       {{
	\count 0 = #1\relax
	\count 2 = #2\relax
	\count 4 = 65536
	\Mess@ge {Before scaling, count 0 = \the \count 0 \space and
			count 2 = \the \count 2}%
	\ifnum	\count 0 > 32767 
	\then	\divide \count 0 by 4
		\divide \count 4 by 4
	\else	\ifnum	\count 0 < -32767
		\then	\divide \count 0 by 4
			\divide \count 4 by 4
		\else
		\fi
	\fi
	\ifnum	\count 2 > 32767 
	\then	\divide \count 2 by 4
		\divide \count 4 by 4
	\else	\ifnum	\count 2 < -32767
		\then	\divide \count 2 by 4
			\divide \count 4 by 4
		\else
		\fi
	\fi
	\multiply \count 0 by \count 2
	\divide \count 0 by \count 4
	\xdef \product {#1 = \the \count 0 \internal@nits}%
	\aftergroup \product
       }}

\def\r@duce{\ifdim\dimen0 > 90\r@dian \then   
		\multiply\dimen0 by -1
		\advance\dimen0 by 180\r@dian
		\r@duce
	    \else \ifdim\dimen0 < -90\r@dian \then  
		\advance\dimen0 by 360\r@dian
		\r@duce
		\fi
	    \fi}

\def\Sine#1%
       {{%
	\dimen 0 = #1 \r@dian
	\r@duce
	\ifdim\dimen0 = -90\r@dian \then
	   \dimen4 = -1\r@dian
	   \c@mputefalse
	\fi
	\ifdim\dimen0 = 90\r@dian \then
	   \dimen4 = 1\r@dian
	   \c@mputefalse
	\fi
	\ifdim\dimen0 = 0\r@dian \then
	   \dimen4 = 0\r@dian
	   \c@mputefalse
	\fi
	\ifc@mpute \then
		\divide\dimen0 by 180
		\dimen0=3.141592654\dimen0
		\dimen 2 = 3.1415926535897963\r@dian 
		\divide\dimen 2 by 2 
		\Mess@ge {Sin: calculating Sin of \nodimen 0}%
		\count 0 = 1 
		\dimen 2 = 1 \r@dian 
		\dimen 4 = 0 \r@dian 
		\loop
			\ifnum	\dimen 2 = 0 
			\then	\stillc@nvergingfalse 
			\else	\stillc@nvergingtrue
			\fi
			\ifstillc@nverging 
			\then	\term {\count 0} {\dimen 0} {\dimen 2}%
				\advance \count 0 by 2
				\count 2 = \count 0
				\divide \count 2 by 2
				\ifodd	\count 2 
				\then	\advance \dimen 4 by \dimen 2
				\else	\advance \dimen 4 by -\dimen 2
				\fi
		\repeat
	\fi		
			\xdef \sine {\nodimen 4}%
       }}

\def\Cosine#1{\ifx\sine\UnDefined\edef\Savesine{\relax}\else
		             \edef\Savesine{\sine}\fi
	{\dimen0=#1\r@dian\advance\dimen0 by 90\r@dian
	 \Sine{\nodimen 0}
	 \xdef\cosine{\sine}
	 \xdef\sine{\Savesine}}}	      

\def\psdraft{
	\def\@psdraft{0}
}
\def\psfull{
	\def\@psdraft{100}
}

\psfull

\newif\if@scalefirst
\def\psscalefirst{\@scalefirsttrue}
\def\psrotatefirst{\@scalefirstfalse}
\psrotatefirst

\newif\if@draftbox
\def\psnodraftbox{
	\@draftboxfalse
}
\def\psdraftbox{
	\@draftboxtrue
}
\@draftboxtrue

\newif\if@prologfile
\newif\if@postlogfile
\def\pssilent{
	\@noisyfalse
}
\def\psnoisy{
	\@noisytrue
}
\psnoisy
\newif\if@bbllx
\newif\if@bblly
\newif\if@bburx
\newif\if@bbury
\newif\if@height
\newif\if@width
\newif\if@rheight
\newif\if@rwidth
\newif\if@angle
\newif\if@clip
\newif\if@verbose
\newif\if@scale
\def\@p@@sclip#1{\@cliptrue}

\newif\if@decmpr

\def\@p@@sfigure#1{\def\@p@sfile{null}\def\@p@sbbfile{null}
	        \openin1=#1.bb
		\ifeof1\closein1
	        	\openin1=\figurepath#1.bb
			\ifeof1\closein1
			        \openin1=#1
				\ifeof1\closein1%
				       \openin1=\figurepath#1
					\ifeof1
					   \ps@typeout{Error, File #1 not found}
						\if@bbllx\if@bblly
				   		\if@bburx\if@bbury
			      				\def\@p@sfile{#1}%
			      				\def\@p@sbbfile{#1}%
							\@decmprfalse
				  	   	\fi\fi\fi\fi
					\else\closein1
				    		\def\@p@sfile{\figurepath#1}%
				    		\def\@p@sbbfile{\figurepath#1}%
						\@decmprfalse
	                       		\fi%
			 	\else\closein1%
					\def\@p@sfile{#1}
					\def\@p@sbbfile{#1}
					\@decmprfalse
			 	\fi
			\else
				\def\@p@sfile{\figurepath#1}
				\def\@p@sbbfile{\figurepath#1.bb}
				\@decmprtrue
			\fi
		\else
			\def\@p@sfile{#1}
			\def\@p@sbbfile{#1.bb}
			\@decmprtrue
		\fi}

\def\@p@@sfile#1{\@p@@sfigure{#1}}

\def\@p@@sbbllx#1{
		\@bbllxtrue
		\dimen100=#1
		\edef\@p@sbbllx{\number\dimen100}
}
\def\@p@@sbblly#1{
		\@bbllytrue
		\dimen100=#1
		\edef\@p@sbblly{\number\dimen100}
}
\def\@p@@sbburx#1{
		\@bburxtrue
		\dimen100=#1
		\edef\@p@sbburx{\number\dimen100}
}
\def\@p@@sbbury#1{
		\@bburytrue
		\dimen100=#1
		\edef\@p@sbbury{\number\dimen100}
}
\def\@p@@sheight#1{
		\@heighttrue
		\dimen100=#1
   		\edef\@p@sheight{\number\dimen100}
}
\def\@p@@swidth#1{
		\@widthtrue
		\dimen100=#1
		\edef\@p@swidth{\number\dimen100}
}
\def\@p@@srheight#1{
		\@rheighttrue
		\dimen100=#1
		\edef\@p@srheight{\number\dimen100}
}
\def\@p@@srwidth#1{
		\@rwidthtrue
		\dimen100=#1
		\edef\@p@srwidth{\number\dimen100}
}
\def\@p@@sangle#1{
		\@angletrue
		\edef\@p@sangle{#1} 
}
\def\@p@@srotate#1{\@p@@sangle{-#1}}
\def\@p@@sscale#1{
		\@scaletrue
		\edef\@p@sscale{#1}
}
\def\@p@@ssilent#1{ 
		\@verbosefalse
}
\def\@p@@sprolog#1{\@prologfiletrue\def\@prologfileval{#1}}
\def\@p@@spostlog#1{\@postlogfiletrue\def\@postlogfileval{#1}}
\def\@cs@name#1{\csname #1\endcsname}
\def\@setparms#1=#2,{\@cs@name{@p@@s#1}{#2}}
\def\ps@init@parms{
		\@bbllxfalse \@bbllyfalse
		\@bburxfalse \@bburyfalse
		\@heightfalse \@widthfalse
		\@rheightfalse \@rwidthfalse
		\@scalefalse
		\def\@p@sbbllx{}\def\@p@sbblly{}
		\def\@p@sbburx{}\def\@p@sbbury{}
		\def\@p@sheight{}\def\@p@swidth{}
		\def\@p@srheight{}\def\@p@srwidth{}
		\def\@p@sangle{0}
		\def\@p@sfile{} \def\@p@sbbfile{}
		\def\@p@scost{10}
		\def\@sc{}
		\@prologfilefalse
		\@postlogfilefalse
		\@clipfalse
		\if@noisy
			\@verbosetrue
		\else
			\@verbosefalse
		\fi
}
\def\parse@ps@parms#1{
	 	\@psdo\@psfiga:=#1\do
		   {\expandafter\@setparms\@psfiga,}}
\newif\ifno@bb
\def\bb@missing{
	\if@verbose{
		\ps@typeout{psfig: searching \@p@sbbfile \space  for bounding box}
	}\fi
	\no@bbtrue
	\epsf@getbb{\@p@sbbfile}
        \ifno@bb \else \bb@cull\epsf@llx\epsf@lly\epsf@urx\epsf@ury\fi
}	
\def\bb@cull#1#2#3#4{
	\dimen100=#1 bp\edef\@p@sbbllx{\number\dimen100}
	\dimen100=#2 bp\edef\@p@sbblly{\number\dimen100}
	\dimen100=#3 bp\edef\@p@sbburx{\number\dimen100}
	\dimen100=#4 bp\edef\@p@sbbury{\number\dimen100}
	\no@bbfalse
}
\newdimen\p@intvaluex
\newdimen\p@intvaluey
\def\rotate@#1#2{{\dimen0=#1 sp\dimen1=#2 sp
		  \global\p@intvaluex=\cosine\dimen0
		  \dimen3=\sine\dimen1
		  \global\advance\p@intvaluex by -\dimen3
		  \global\p@intvaluey=\sine\dimen0
		  \dimen3=\cosine\dimen1
		  \global\advance\p@intvaluey by \dimen3
		  }}
\def\compute@bb{
		\no@bbfalse
		\if@bbllx \else \no@bbtrue \fi
		\if@bblly \else \no@bbtrue \fi
		\if@bburx \else \no@bbtrue \fi
		\if@bbury \else \no@bbtrue \fi
		\ifno@bb \bb@missing \fi
		\ifno@bb \ps@typeout{FATAL ERROR: no bb supplied or found}
			\no-bb-error
		\fi
		\count203=\@p@sbburx
		\count204=\@p@sbbury
		\advance\count203 by -\@p@sbbllx
		\advance\count204 by -\@p@sbblly
		\edef\ps@bbw{\number\count203}
		\edef\ps@bbh{\number\count204}
		\if@angle 
			\Sine{\@p@sangle}\Cosine{\@p@sangle}
	        	{\dimen100=\maxdimen\xdef\r@p@sbbllx{\number\dimen100}
					    \xdef\r@p@sbblly{\number\dimen100}
			                    \xdef\r@p@sbburx{-\number\dimen100}
					    \xdef\r@p@sbbury{-\number\dimen100}}
                        \def\minmaxtest{
			   \ifnum\number\p@intvaluex<\r@p@sbbllx
			      \xdef\r@p@sbbllx{\number\p@intvaluex}\fi
			   \ifnum\number\p@intvaluex>\r@p@sbburx
			      \xdef\r@p@sbburx{\number\p@intvaluex}\fi
			   \ifnum\number\p@intvaluey<\r@p@sbblly
			      \xdef\r@p@sbblly{\number\p@intvaluey}\fi
			   \ifnum\number\p@intvaluey>\r@p@sbbury
			      \xdef\r@p@sbbury{\number\p@intvaluey}\fi
			   }
			\rotate@{\@p@sbbllx}{\@p@sbblly}
			\minmaxtest
			\rotate@{\@p@sbbllx}{\@p@sbbury}
			\minmaxtest
			\rotate@{\@p@sbburx}{\@p@sbblly}
			\minmaxtest
			\rotate@{\@p@sbburx}{\@p@sbbury}
			\minmaxtest
			\edef\@p@sbbllx{\r@p@sbbllx}\edef\@p@sbblly{\r@p@sbblly}
			\edef\@p@sbburx{\r@p@sbburx}\edef\@p@sbbury{\r@p@sbbury}
		\fi
		\count203=\@p@sbburx
		\count204=\@p@sbbury
		\advance\count203 by -\@p@sbbllx
		\advance\count204 by -\@p@sbblly
		\edef\@bbw{\number\count203}
		\edef\@bbh{\number\count204}
}
\def\in@hundreds#1#2#3{\count240=#2 \count241=#3
		     \count100=\count240	
		     \divide\count100 by \count241
		     \count101=\count100
		     \multiply\count101 by \count241
		     \advance\count240 by -\count101
		     \multiply\count240 by 10
		     \count101=\count240	
		     \divide\count101 by \count241
		     \count102=\count101
		     \multiply\count102 by \count241
		     \advance\count240 by -\count102
		     \multiply\count240 by 10
		     \count102=\count240	
		     \divide\count102 by \count241
		     \count200=#1\count205=0
		     \count201=\count200
			\multiply\count201 by \count100
		 	\advance\count205 by \count201
		     \count201=\count200
			\divide\count201 by 10
			\multiply\count201 by \count101
			\advance\count205 by \count201
		     \count201=\count200
			\divide\count201 by 100
			\multiply\count201 by \count102
			\advance\count205 by \count201
		     \edef\@result{\number\count205}
}
\def\ps@scaleinhundreds#1{
		\in@hundreds{#1}{\@p@sscale}{100}
		\edef#1{\@result}
}
\def\compute@wfromh{
		\in@hundreds{\@p@sheight}{\@bbw}{\@bbh}
		\edef\@p@swidth{\@result}
}
\def\compute@hfromw{
	        \in@hundreds{\@p@swidth}{\@bbh}{\@bbw}
		\edef\@p@sheight{\@result}
}
\def\compute@handw{
		\if@height 
			\if@width
			\else
				\compute@wfromh
			\fi
		\else 
			\if@width
				\compute@hfromw
			\else
				\edef\@p@sheight{\@bbh}
				\edef\@p@swidth{\@bbw}
			\fi
		\fi
}
\def\compute@resv{
		\if@rheight \else \edef\@p@srheight{\@p@sheight} \fi
		\if@rwidth \else \edef\@p@srwidth{\@p@swidth} \fi
}
\def\compute@sizes{
	\compute@bb
	\if@scalefirst\if@angle
	\if@width
	   \in@hundreds{\@p@swidth}{\@bbw}{\ps@bbw}
	   \edef\@p@swidth{\@result}
	\fi
	\if@height
	   \in@hundreds{\@p@sheight}{\@bbh}{\ps@bbh}
	   \edef\@p@sheight{\@result}
	\fi
	\fi\fi
	\compute@handw
	\compute@resv
	\if@scale
	   \if@verbose
	      \ps@typeout{(scaling by \@p@sscale)}%
	   \fi
	   \ps@scaleinhundreds{\@p@swidth}%
	   \ps@scaleinhundreds{\@p@sheight}%
	   \ps@scaleinhundreds{\@p@srwidth}%
	   \ps@scaleinhundreds{\@p@srheight}%
	\fi
}

\def\psfig#1{\vbox {
	\ps@init@parms
	\parse@ps@parms{#1}
	\compute@sizes
	\ifnum\@p@scost<\@psdraft{
		\special{ps::[begin] 	\@p@swidth \space \@p@sheight \space
				\@p@sbbllx \space \@p@sbblly \space
				\@p@sbburx \space \@p@sbbury \space
				startTexFig \space }
		\if@angle
			\special {ps:: \@p@sangle \space rotate \space} 
		\fi
		\if@clip{
			\if@verbose{
				\ps@typeout{(clip)}
			}\fi
			\special{ps:: doclip \space }
		}\fi
		\if@prologfile
		    \special{ps: plotfile \@prologfileval \space } \fi
		\if@decmpr{
			\if@verbose{
				\ps@typeout{psfig: including \@p@sfile.Z \space }
			}\fi
			\special{ps: plotfile "`zcat \@p@sfile.Z" \space }
		}\else{
			\if@verbose{
				\ps@typeout{psfig: including \@p@sfile \space }
			}\fi
			\special{ps: plotfile \@p@sfile \space }
		}\fi
		\if@postlogfile
		    \special{ps: plotfile \@postlogfileval \space } \fi
		\special{ps::[end] endTexFig \space }
		\vbox to \@p@srheight true sp{
			\hbox to \@p@srwidth true sp{
				\hss
			}
		\vss
		}
	}\else{
		\if@draftbox{		
			\hbox{\frame{\vbox to \@p@srheight true sp{
			\vss
			\hbox to \@p@srwidth true sp{ \hss \@p@sfile \hss }
			\vss
			}}}
		}\else{
			\vbox to \@p@srheight true sp{
			\vss
			\hbox to \@p@srwidth true sp{\hss}
			\vss
			}
		}\fi

	}\fi
}}
\psfigRestoreAt
\let\@=\LaTeXAtSign

\def\etal{\rm et al.}
\def\p{$\pm$}
\def\simlt{\lower.5ex\hbox{$\; \buildrel < \over \sim \;$}}
\def\msol{M$_{\sun}$}

\begin{document}

\thesaurus{
              (08.12.2;  
               08.12.1)  
     } 

\title{Optical Spectroscopy of DENIS Mini-survey Brown Dwarf Candidates\thanks{
Based on observations made at the Anglo-Australian Telescope, Siding Spring.}}

\author{C.G. Tinney\inst{1}, 
        X. Delfosse\inst{2,3},   
        T. Forveille\inst{3}, 
        F. Allard\inst{4,5}.} 

\offprints{C.Tinney, cgt@aaoepp.aao.gov.au}
\institute{
              Anglo-Australian Observatory, 
              PO Box 296, 
              Epping. N.S.W. 2121. 
              Australia 
\and
              Observatoire de Gen\`eve,
              CH-1290 Sauverny,
              Switzerland
\and
              Observatoire de Grenoble,
              Domaine Universitaire de S$^{\mathrm t}$ Martin d'H\`eres,
              F-38041 Grenoble,
              France
\and
              Centre de Recherche Astronomique de Lyon (UMR 142 CNRS), 
              Ecole Normale Superieure, 
              69364 Lyon Cedex 07, 
              France,
\and
              Department of Physics, 
              Wichita State University, Wichita, 
              KS 67260-0032, USA,
}

\date{Received ; accepted}

\maketitle

\markboth{Tinney \etal: Spectroscopy of DENIS Brown Dwarf Candidates}{}

\begin{abstract}
We present optical (6500-9200\AA) spectroscopy of eight cool dwarfs detected
in a 231 square degree ``Mini-survey'' of the Deep NEar Infrared Survey
(DENIS) data. We are able to confirm that the spectral types derived
from the Mini-survey infrared spectroscopy are meaningful.
We provide a spectral sequence which extends beyond the M-dwarf range
and into the proposed ``L'' class of dwarfs. The dominant spectral
features in the optical for these L-type dwarfs are resonance lines of
Cs\,I and molecular band heads of CrH and FeH. The other dominant
feature in these L-type spectra is a broad 600\,\AA\ absorption dip centered
on 7700\,\AA, which we identify with extremely strong (equivalent width $\sim$
several hundred \AA) absorption associated with the 7664,7698\,\AA\ resonance
doublet of K\,I. We find that model atmospheres which include the effects
of molecular condensation without dust opacity (to simulate rapid gravitational
settling of dust grains) produce significantly better agreement with
observed optical spectra for L-type dwarfs, than models including dust opacity.
This suggests gravitational settling of dust grains  plays an important
role in L-dwarf photospheres. The extreme strength of the K\,I resonance 
doublet, and disappearance of TiO and VO, and the consequent dominance of
CrH and FeH in L-dwarf spectra offer considerable prospects as sensitive
effective temperature diagnostics, even at low spectral resolution.

\keywords{stars: low-mass,brown dwarfs - late-type
           }

\end{abstract}

\section{Introduction}

The history of the study of very-low mass (VLM) stars and brown dwarfs has
shown again and again that when new technologies are implemented, new
objects with previously unseen properties are discovered. 
Examples include the use of wide-field photographic surveys
and digital scanning machines to discover the first VLM stars (Luyten 1979;
Reid \& Gilmore 1981; Probst \& Liebert 1983; Bessell 1991; Irwin \etal\ 1991);
the use of infrared spectroscopy to discover the importance of H$_2$O
absorption in VLM stars (Berriman \& Reid 1987); and, the use of infrared
imaging, adaptive optics and coronography to discover the proto-typical cold brown dwarf
Gl\,229B, which further confirmed the importance of CH$_4$ in cold brown dwarfs
(Nakajima \etal\ 1995; Oppenheimer \etal\ 1995).
The next major breakthrough will be
the identification of significant numbers of brown dwarfs by
the coming generation of infrared all-sky surveys -- in particular
DENIS (Epchtein 1997) and 2MASS (Skrutskie \etal\ 1997). \\

DENIS will be a complete near infrared survey of the southern sky
(Epchtein \etal\ 1994; Epchtein 1997) in the I, J and K$^\prime$
bands, to approximate 3-$\sigma$ limits of I=18, J=16, and K=13.5. The
products of this survey will be databases of calibrated images,
extended sources, and small objects. The survey started in January
1996 and is expected to be completed within five years.  We have
carried out a ``Mini-survey'' with infrared spectroscopic follow up on
the very low-mass (VLM) star and brown dwarf candidates contained in
$\approx$1\% of the DENIS survey data (Delfosse \etal\ 1998, 1997).  
The image data from the high latitude part
($|b_{II}|>20$-30\degr) of 47 survey strips, were processed and used to
identify a sample of objects for which infrared H- and K-band
spectroscopy was carried out in order to estimate
luminosities/temperatures. In this paper 
we present optical spectroscopy for a sample of cool dwarfs identified in
a 231 square degree ``Mini-survey'' of the data from the DEep Near-Infrared 
Survey (DENIS), and discuss the significant features these spectra reveal. 

\section{Optical Spectroscopy}

Optical spectroscopy of a sample of the Mini-survey sources was
obtained with the AAT on 1997 June 7-9 (UT), using the RGO
Spectrograph with TEK 1K CCD\#2. Observations were made using a 270R
grating in blaze-to-camera mode with a 1.7\arcsec\ slit, providing a
resolution of 7\,\AA\ and a wavelength coverage of 6425--9800\,\AA.
The observations of both our DENIS Mini-survey targets and several
comparison VLM stars are summarised in Table \ref{table}. Total
exposure times of more than 1800s, were obtained as multiple half-hour
exposures.  Finding charts for the DENIS objects can be found in
Delfosse \etal\ (1998). The data were processed using standard
techniques within the FIGARO data reduction package (Shortridge 1993).
In particular the data were: bias subtracted; trimmed; flat fielded;
straightened; sky-subtracted; cleaned of cosmic rays by hand;
optimally extracted; and, flux calibrated using observations of HST
standard stars (Turnshek \etal\ 1990). Photon counting errors were
propagated through the reduction, so that as a last step, multiple
exposures (after normalising to the same mean level in the wavelength
range 8300-8400\AA) were combined to produce a final weighted mean
spectrum. Seeing ranged from 1\arcsec\ to 2\arcsec\ over the run.

\subsection{The Spectra \label{spectra}}

\begin{table*}
{\caption{Optical Spectroscopy. \label{table}}
   \begin{tabular}{lccccccc}
Object                    & Position$^a$          & Exp. & I           & I--J      & EW(\AA)$^b$ & C$^c$ & Ref$^d$ \\
                          & (J2000.0)             &  (s) &             &           & Cs\,I       &\\[5pt]
DENIS-P\,J0020$-$4414     & 00:20:59.4 $-$44:14:43& 3600 & 18.32\p0.16 & 3.35\p0.17& -&2.0 &1\\ 
DENIS-P\,J0021$-$4244     & 00:21:05.7 $-$42:44:50& 1800 & 16.83\p0.05 & 3.30\p0.05& 0.9\p0.3&3.5 &1\\ 
DENIS-P\,J0205$-$1159     & 02:05:29.0 $-$11:59:25& 5400 & 18.30\p0.24 & 3.67\p0.25& 7.5\p1.0&7.0 &1\\ 
DENIS-P\,J1058$-$1548     & 10:58:46.5 $-$15:48:00& 3600 & 17.80\p0.17 & 3.72\p0.17& 2.6\p0.5&5.0 &1\\ 
DENIS-P\,J1228$-$1547     & 12:28:13.8 $-$15:47:11& 1800 & 18.19\p0.27 & 3.76\p0.27& 5.8\p1.0&6.0 &1\\ 
DENIS-P\,J2040$-$3245     & 20:40:06.2 $-$32:45:24& 3600 & 17.86\p0.16 & 2.97\p0.17& -&1.0 &1\\ 
DENIS-P\,J2052$-$5512     & 20:52:55.0 $-$55:12:03& 3600 & 17.52\p0.13 & 2.70\p0.15& -&0.0 &1\\ 
DENIS-P\,J2146$-$2153     & 21:46:10.6 $-$21:53:09& 3600 & 18.40\p0.27 & 2.98\p0.29& -&3.0 &1\\[5pt] 
BRI\,0021-0214            & 00:24:24.6 $-$01:58:22& 1800 & 15.07\p0.03 & 3.17\p0.05& -&4.0 &2\\
VB\,10/LHS~474            & 19:16:57.9 $+$05:09:10&  600 & 12.80       & 2.90      & -&1.5 &3\\
GRH\,2208                 & 22:10:50.0 $-$19:52:13& 1800 &   -         &  -        & -&0.5 &\\
  \end{tabular}}
\raggedright
\noindent
\vskip 10pt
Notes : \\
$a$ - Positions for the DENIS objects are from Delfosse \etal\ 1998, 
those for the remainder are from Tinney \etal\ 1995
and Tinney 1996.  The DENIS-P prefix indicates that these are
provisional DENIS objects, which have not been produced by the final
DENIS catalogue pipeline.\\
$b$ - Psuedo-continuum defined by polynomial fit to the wavelength range 8430-8560\AA. Blank
      entries imply unmeasurable equivalent widths $\simlt 0.5$\AA.\\
$c$ - Constant offset applied to the spectrum in Fig. \ref{optical}.\\
$d$ - References for photometry: (1) Delfosse \etal\ 1998, (2) Tinney \etal\ 1993, (3) Leggett 1992.\\
 \end{table*}

\begin{figure*}
\psfig{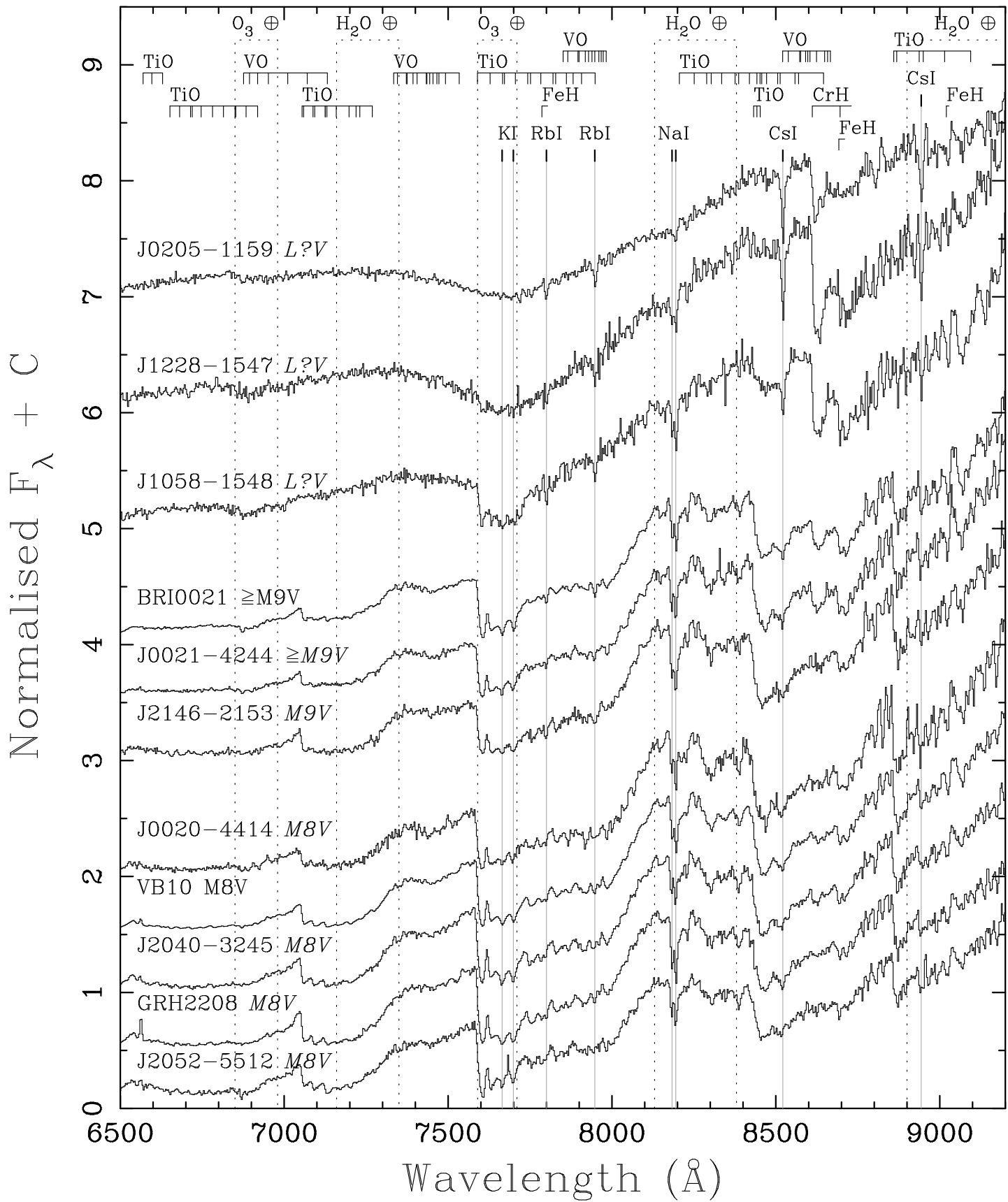}
\caption{Optical Spectra of DENIS Mini-survey sources. The spectra
shown have been normalised in the range 8700-8800\AA, and are offset
by the constants (C) provided in Table \ref{table}. The spectral types
provided and the absorption features marked are discussed in
Sect.\ref{spectra} of the text.}
\label{optical}
\end{figure*}

The resulting spectra are shown in Fig. \ref{optical}.
The spectra were not corrected for the effects of atmospheric
absorption, so the locations of significant terrestrial absorption 
are also indicated. Also shown are the wavelengths of atomic
and molecular absorption features commonly seen in very late
M-dwarf spectra. These are discussed in more detail
by Tinney \& Reid (1998).
Preliminary versions of the coolest of these spectra
were presented by Tinney \etal\ (1997).\\

The spectra shown in Fig. \ref{optical} clearly fall into two
categories: the lower 8 objects are very late M-dwarfs, with prominent
TiO, VO, Na and K features; while the upper three objects are also low
temperature objects, they show either weak, or totally absent, TiO and
VO absorption. Infrared spectroscopy (Delfosse \etal\ 1998, 1997) 
indicates that they are all considerably lower in effective
temperature than M-type dwarfs. The detection of Li in DENIS-P\,J1228-1547 
(Tinney \etal\ 1997; Mart\'\i n \etal\ 1997), indicates a mass of less
than 0.065\msol, and definite brown dwarf status.
The weakness of TiO and VO in their
spectra is understood to be due to the condensation of dust in their
photospheres. This has two main effects: (1) it creates a
``greenhouse'' effect which warms the upper photosphere, weakening
the bands of H$_2$O, TiO and VO (Tsuji \etal\ 1996a; Jones \& Tsuji 1997) ;
and (2) below T$_{\rm eff} \sim 2600$\,K dust condensates
-- in particular perovskite (CaTiO$_3$) and solid VO, containing Ti and V 
-- will begin to condense out of the photosphere, therefore depleting it of
TiO and VO (Allard 1998; Allard \etal\ 1997; Sharp \& Huebner 1990). 
This has lead to the suggestion by both Kirkpatrick (1998) that there is a clear need for a new spectral class for
these objects. Although a definitive allocation has yet to be made, we
adopt here the preliminary ``L'' designation for these very low
temperature dwarfs.\\

We have therefore ordered the spectra in apparent order of decreasing
temperature.  In the case of the M-dwarfs this has been done by
comparison with the spectra for BRI\,0021 and VB\,10 (the spectral
types for which are due to Kirkpatrick \etal\ 1997b and
Kirkpatrick \etal\ 1995 respectively) on the typing system
of Kirkpatrick \etal\ (1991) and Kirkpatrick \etal\
(1995). Our estimated spectral types are shown in italics in Fig.
\ref{optical}. These are not based on the detailed least-squares
fitting procedure adopted by Kirkpatrick \etal\ (1991), but by eyeball
comparison of the reference spectra, in particular concentrating on
the strength of the VO bands at 7445 and 7850\,\AA, and the TiO bands
at 7050 and 7600\,\AA. A more quantitative typing was also attempted using the
psuedo-continuum ratios of Mart\'\i n \etal\ (1996) --
in particular their ``PC3'' index, the values for which are shown in
Table \ref{table2}. This produced exactly the same ordering
for the M-dwarfs as the eyeball comparison shown in Fig. 1. However, it produced 
sub-types for VB10 and BRI\,0021-0214 at variance with their standard values by $\approx$ 0.5
sub-types, and systematically high. Given this we have chosen to only provide
estimated spectral types to the nearest unitary sub-type in Fig. 1.\\

In the case of the L-dwarfs, the ordering has
been performed based on the infrared spectroscopy of Delfosse \etal\
(1998). The absence of a well defined typing scheme for the L-dwarfs
leads us to leave the subtypes as undetermined (``?'' in the figure),
though work on this front is in hand (Kirkpatrick \etal\ 1998b).\\

\section{Discussion}

\subsection{The M-dwarf Spectra}

The latest DENIS M-dwarf (DENIS-P\,J0021, $\ge$M9V) shows an almost
identical spectrum to the very late M-dwarf BRI\,0021-0214, which
itself shows one of the latest M spectra known for a field dwarf
(only 2MASSP\,J0345 is
known to have a later M-type spectrum; Kirkpatrick \etal\ 1997a). 
DENIS-P\,J2146 shows a slightly earlier spectral
type of M9V. The remaining DENIS M-dwarfs all show types of
M8V. GRH\,2208 was discovered by Gilmore \etal\ (1985), who
classified it as a possible VLM star. Its M8V spectral type and large
proper motion (Tinney 1996) confirm this.\\

Of the M-dwarfs observed only a few show evidence for H$\alpha$
emission: GRH\,2208 has emission at an equivalent width of
$\approx$5\,\AA, while VB10, DENIS-P\,J2052, DENIS-P\,J0020 and
DENIS-P\,J0020 show evidence for emission at the $\sim$1\,\AA\ level.

\subsection{The L-dwarf Spectra}

As discussed above, the most prominent feature of the L-dwarf spectra
is weak, or non-existent, TiO and VO absorption. Also obvious is
the strong resonance line of Cs\,I at 8521.4\,\AA\ (equivalent widths
for which are provided in Table 1) and a pair of
strong bandheads at 8610 and 8700\,\AA. These latter bandheads
have recently been identified by Kirkpatrick \etal\ (1998a) as being
due to CrH bandheads at 8611\,\AA\ and 8696\,\AA\ (Pearse \& Gaydon 1976), and
an FeH bandhead at  8692\,\AA\ (Phillips \etal\ 1987). \\

It is interesting to note, however, 
that once one has identified these bands
in the spectra of L-dwarfs, it becomes obvious that they are also present
(though somewhat masked by TiO and VO) in the spectra of the latest 
M-dwarfs. All
of the dwarfs of M9V or later show the same CrH/FeH bandheads.
It is also interesting to see that the ordering provided by the
the infrared spectra for the L-dwarfs in combination with
the M spectral types, shows a clear sequence of
decreasing TiO and VO band strength with decreasing temperature.
In particular, DENIS-P\,J1058  shows weaker TiO at 8400 and
7600\,\AA\ than BRI\,0021, while DENIS-P\,J1228 shows only
hints of TiO at 8400\,\AA\ and DENIS-P\,J0205 shows no evidence
for TiO or VO at all.\\

If we assume that the temperature ordering we have adopted is
reasonable, we can then see that the CrH/FeH bands appear 
{\em weaker} in the coolest object (DENIS-P\,J0205), than in the
two earlier type L-dwarfs. This is not necessarily contrary to what one might
expect. Molecular equilibrium
calculations by Sharp \& Huebner (1990), have shown that at 
plasma temperatures of $\approx$1420\,K solid Fe will begin to condense, 
and similarly that at plasma temperatures of $\approx$1250\,K Cr$_2$O$_3$ 
will begin to condense, depleting the
photosphere of CrH. In other words, we can expect CrH and FeH to become
prominent when TiO and VO are depleted by grain formation, but that
they themselves will weaken at a somewhat lower effective temperature. 
Such an interpretation of the observed behavior of these bands is
supported by the fact that {\em no} evidence for CrH or FeH is
seen in the optical spectrum of the T$_{\rm eff}\approx 900$\,K
brown dwarf Gl\,229B (Oppenheimer \etal\ 1998; Schultz \etal\ 1998), 
which shows only Cs\,I, H$_2$O and possibly CH$_4$ features in the optical.\\

This ``on-again-off-again'' behavior offers tantalising possibilities for
future brown dwarf studies. The difficulty of making
model atmospheres deal with the complex opacities needed to match
observed VLM spectra has severely hampered progress in their understanding
(Allard \etal\ 1997).
In particular, it has proved difficult in the extreme to estimate precise
and robust effective temperatures below $\approx$2500\,K. However, the appearance and 
disappearance of bands like CrH over a fairly narrow range of temperatures
offers the possibility of a temperature diagnostic which is both easily observed, 
and which should be accessible to photospheric models including the effects of
dust condensation. \\

Care must be taken however to distinguish between the plasma temperature
at which a species will condense, from a dwarf's overall effective temperature.
For example, the atmospheric models discussed in section \ref{models}, 
show that solid Fe will begin to condense in objects with T$_{\rm eff}$ considerably
hotter than 1500\,K. However, this does not affect the formation of FeH bands
at T$_{\rm eff}=1500$\,K, because FeH opacities lead to band formation 
deep in the photosphere, where FeH does {\em not} condense at this effective 
temperature. Similar behaviour
may occur for CrH. At present we can only point out that there does appear
to be a considerable change in the observed CrH/FeH band strengths between 
T$_{\rm eff}$=2000 and 1000\,K, and that these offer the prospect of a
modellable and robust temperature sensitive indicator.\\

\begin{figure*}
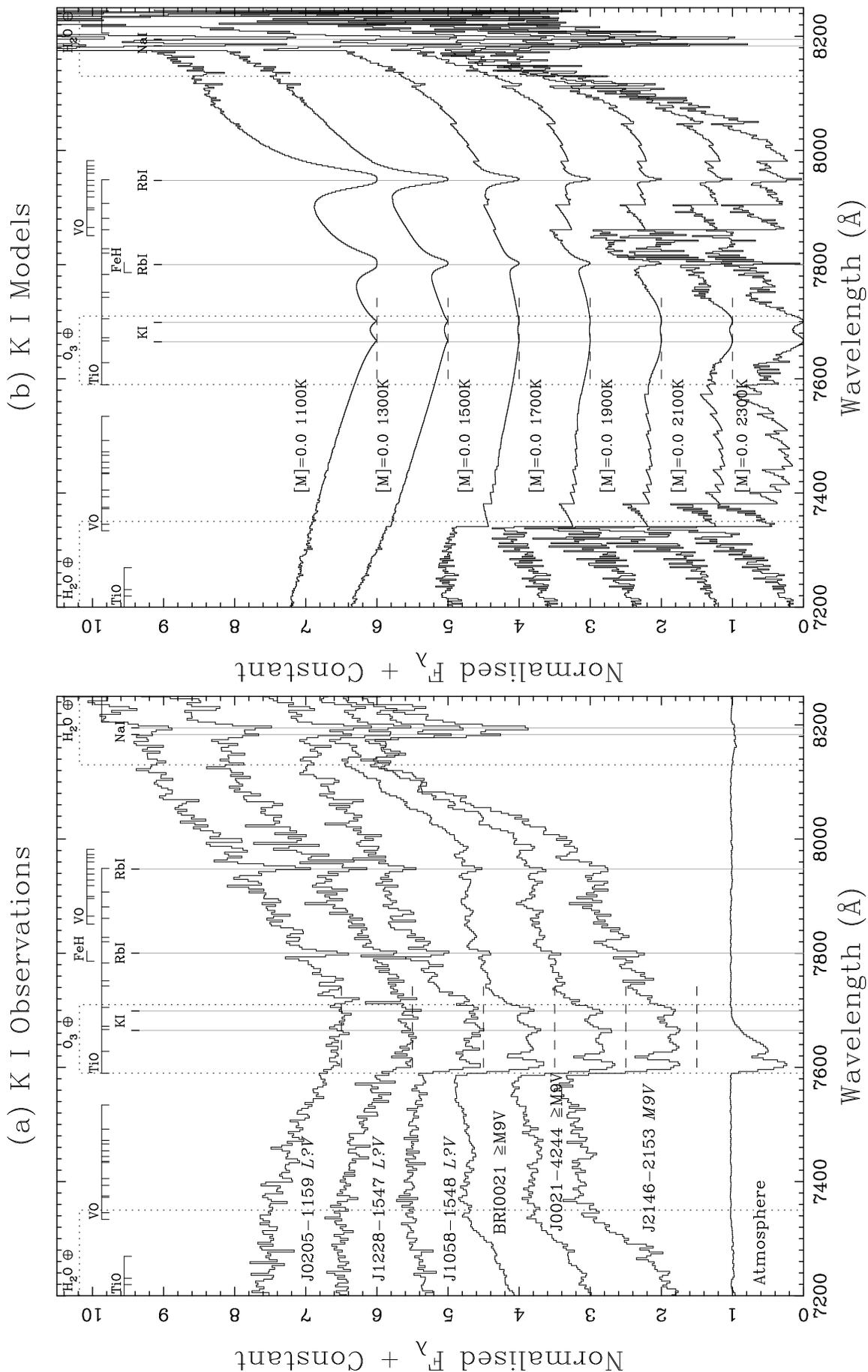

\centerline{\psfig{height=12cm,file=plot_ki_ames_1.ps,rotate=270}}
\vskip 5pt
\centerline{\psfig{height=12cm,file=plot_ki_1.ps,rotate=270}}
\caption{Region of the K\,I Doublet. {\em Left Panel} - expanded
versions of the data for the coolest objects in Fig. \ref{optical},
in which the spectra have been normalised in the range 7275-7325\AA,
and offset. The zero-level for each spectrum is indicated by the
dashed line between 7600 and 7750\,\AA. Also shown is a normalised
standard star spectrum to indicate the location of terrestrial O$_3$
and H$_2$O absorption.  {\em Right Panel} - ``NGAMES-Cond'' atmospheric
models for solar metallicities, log$g$=5.0 at the indicated
temperatures.} 
\label{ki}
\end{figure*}

There is one further obvious feature in the L-dwarf spectra which has
not been discussed. This is the pronounced broad dip seen in
DENIS-P\,J1228 and DENIS-P\,J0205 centered on 7700\,\AA.  Some evidence
for this feature is also seen in DENIS-P\,J1058, though it is masked
by the overlying stellar TiO and terrestrial O$_3$ absorption. Fig.
\ref{ki}(a) shows an expanded version of Fig. \ref{optical} in this
wavelength range for the latest six dwarfs observed.  We see a clear
depression of the L-dwarf ``continuum'' in the neighbourhood of the K\,I lines
near 7700\,\AA\ -- in fact over the wavelength
range 7600-7750\,\AA\ we detect essentially zero flux.\\

Mart\'\i n \etal\ (1997) have discussed the core of this feature, but
the limited spectral coverage of their high resolution echelle orders
prevented them from appreciating its full extent. They suggested that
it is due to an extreme broadening of the 7664,7698\,\AA\ K\,I
doublet.  At first glance, such an interpretation seems hard to believe,
given the huge extent of the absorption seen in our L-dwarf spectra.  It
requires that the absorption line is not only extremely broad, but
that it be totally saturated over a wavelength range of 150\AA\ (or
$\lambda/\Delta\lambda \approx 51$), with the line wings extending
over almost 600\,\AA\ (or $\lambda/\Delta\lambda \approx 13$).  The
total equivalent width (EW) of such a line in the two latest L dwarfs presented
is $\sim$300\,\AA 
\footnote{Measured relative to pseudo-continuum points at 7200 and 8200\AA}.  
This is
almost without precedent in stellar spectra, even for the extremely
strong H or He lines seen in white dwarfs.  Moreover, 
while the spectra of slightly earlier M-dwarfs show K\,I, these lines
represent only EW $\sim$ 10-20\,\AA\ absorptions.  If the 7700\,\AA\
dip is due to K\,I then it requires that the photosphere, for a change
of only a few hundred degrees in effective temperature, produces a
massive change in K\,I absorption equivalent width.\\

\subsection{Model Atmospheres including Condensation\label{models}}

\begin{table*}
{\caption{Comparison of Optical and Infrared Spectral Typing \label{table2}}
   \begin{tabular}{lcccc}
Object                    &  I--J     & Infrared Spectrum  & PC3$^c$ & Spectral  \\
                          &           & M$_K$ Estimate$^a$ &     & Type$^b$      \\[5pt]
DENIS-P\,J0205$-$1159     & 3.67\p0.25& 12.1$\pm$0.5	   &11.21& L?\\ 
DENIS-P\,J1228$-$1547     & 3.76\p0.27& 12.0$\pm$0.5	   &6.86 & L?\\ 
DENIS-P\,J1058$-$1548     & 3.72\p0.17& 11.5$\pm$0.5	   &3.26 & L?\\ 
DENIS-P\,J0021$-$4244     & 3.30\p0.05& 11.3$\pm$0.5	   &2.44 & $\ge$M9V\\ 
DENIS-P\,J2146$-$2153     & 2.98\p0.29& 11.4$\pm$0.6	   &2.39 & M9V\\ 
DENIS-P\,J0020$-$4414     & 3.35\p0.17& 11.0$\pm$0.5	   &2.18 & M8V\\ 
DENIS-P\,J2040$-$3245     & 2.97\p0.17& 11.1$\pm$0.5	   &1.66 & M8V\\ 
DENIS-P\,J2052$-$5512     & 2.70\p0.15& 10.7$\pm$2.8	   &1.56 & M8V\\ 
  \end{tabular}}
\raggedright
\noindent
\vskip 10pt
Notes : \\
$a$ - Estimated absolute K magnitudes based on infrared spectral
      characteristics due to Delfosse \etal\ 1998.\\
$b$ - PC3 index defined for spectral typing Mart\'\i n \etal\ 1996. Although
      Mart\'\i n \etal\ do not actually provide the flux scale used to define
      the index, we have used F$_\lambda$.\\
$c$ - Spectral types from Fig. \ref{optical} derived as described
      in Sect.\ref{spectra}.
 \end{table*}

In order to investigate the nature of the 7700\,\AA\ dip, it is necessary
to refer to detailed atmospheric models. The construction of model atmospheres
which include the effects of dust on the equation of state and radiative 
transfer, is a field still in its infancy. We show in Fig. \ref{ki}(b)
preliminary ``NextGen-AMES-Cond'' or ``NGAMES-Cond'' models due to F.Allard \& P.Hauschildt
for a range of temperatures in the region of the K\,I line. These
models were constructed using the full ``Opacity Sampling'' (OS) technique of
the ``NextGen'' models of Allard \etal\ (1997) with the further addition
of; the AMES H$_2$O opacities of Partridge \& Schwenke (1997); and molecular
condensation equilibria using the thermodynamic equlibrium constants of Sharp \& Huebner (1990).
They differ from the ``NGAMES-Dusty'' models of Allard \& Hauschildt 
(Allard 1997; Allard \etal\ 1998; Kirkpatrick \etal\ 1998a) in that they do not
include the effects of dust opacities on radiative transfer.
The ``NGAMES-Dusty'' models have been used by Kirkpatrick \etal\ (1998a)
to study the optical-infrared spectrum of GD\,165B. They found that
the inclusion of dust in the radiative transfer dramatically improved the
agreement of the models to the observed infrared spectrum. However, they
also found that dust opacity resulted in a significant over-prediction of 
TiO and VO band strengths in the optical, because of heating of the upper 
photosphere by the greenhouse effect, which among other things will reduce
the rate of grain formation, and so of TiO and VO depletion. The over-prediction
of these bands by the models is thought to take place
because they do not take into account gravitational settling,
which will actually deplete the upper photosphere of dust.
As we are most interested in examining optical spectra, we have
therefore used the ``NGAMES-Cond'' models to simulate the effects of 
gravitational settling, since they essentially make grains ``disapear'' 
after they have formed.\\

Kirkpatrick \etal\ (1998a) derived an effective temperature of 
1900$\pm$100\,K for GD\,165B. Given the similarity of GD\,165B's spectrum to
the L-dwarfs shown in Fig. \ref{ki}(a) (Tinney \etal\ 1997), we therefore assume 
that their effective temperatures lie somewhere in the range
2000-1600\,K. The NGAMES-Cond models predict that at these effective temperatures,
opacity due to TiO/VO does indeed decrease due to the combined effects of
condensation and dust settling, and that
the lines of the K\,I doublet become {\em very} strong. Fig. \ref{ki} also
shows some disagreement between the models and the data -- in particular,
the models still seem to over-predict absorption due to TiO near 7200\,\AA\ and
VO at 7340-7560 and 7870-7960\,AA, as well as predicting larger equivalent widths
for the resonance lines of Rb\,I than are observed. This is not surprising,
given crude assumption of infinitely rapid settling for all grain
species.  The models also fail to predict
features due to CrH, which is not included in the current opacity line list.
It would clearly, therefore, be premature to draw detailed 
conclusions from these models. However, we can clearly see that they produce a
much better match to observations than the ``NGAMES-Dusty'' presented
by Kirkpatrick \etal\ (1998a -- Fig.4), indicating that gravitational settling
is clearly taking place in the upper photospheres of L-dwarfs. Moreover, they show that the
effects of dust condensation do produce qualitatively the correct
trend for decreasing molecular absorption, and that K\,I is responsible for the  
7700\,\AA\ dip in the ``molecule-poor'' photospheres of L-dwarfs.\\

The reality of dust settling will clearly lie somewhere in between the
rapid settling modelled by the NGAMES-Cond models, and the absence of
settling indicated by the NGAMES-Dusty models. The infrared spectrum of 
Gl\,229B shows both no evidence for dust in its infrared spectrum 
(Oppenheimer \etal\ 1998, Tsuji \etal\ 1996b), {\em and} evidence for dust 
in its optical spectrum (Schultz \etal\ 1998).
Pavlenko \etal\ (1998) have produced models which can reproduce L-dwarf spectra, 
by including an {\em a postiori} power-law opacity component in the optical.
However, such model does not lead to any immediate physical
understanding of the processes which produce such an opacity law.
Clearly simple dust models cannot reproduce the complex behaviour seen in
L-dwarfs. Real understanding will have to await the coming generation of
atmospheric models which include the chemical equilibria of dust formation,
an understanding of the grain size distribution, and the physics of gravitational
settling.\\ 

The relative weakness of other neutral lines in our observed spectra, despite
the strength of the K\,I doublet, is also reasonable. The lines of
Na\,I at 8183 and 8194\,\AA\ are not resonance lines like the K\,I
doublet -- they would therefore be expected to be significantly
weaker in a very cool, dense photosphere. The lines of Rb\,I at 7800
and 7947\,\AA\ and Cs\,I at 8521 and 8943\,\AA\ are resonance lines
and have similar line strengths to the K\,I doublet (Corliss \& Buzman
1962), {\em but} have abundances $\sim 3500$ and 7000 times
(respectively) lower than K (Allen 1976).  So again, it is reasonable
to expect them to be significantly weaker than the K\,I doublet. The
lines of the Na\,I doublet at 5890,5896\,\AA (which lie outside the
wavelength coverage of our spectra), on the other hand {\em are}
resonance transitions with similar line strengths to the K\,I doublet,
and indeed the ``NGAMES-Cond'' models predict these lines to be
even stronger than K\,I. The huge strength of these high abundance
resonance lines offers exciting possibilities for the
estimation of photospheric properties (in particular effective
temperatures) with quite low resolution spectra, once the details
of grain growth and settling are understood.

\subsection{A Comparison of Infrared and Optical Spectral Typing}

In order to refine the sample produced by pure colour selection
on the DENIS survey data, Delfosse \etal\ (1998) obtained infrared
spectra for {\em all} the DENIS Mini-survey VLM and brown dwarf
candidates. These were then used to define H$_2$O indices which
were used to ``type'' spectra in the infrared. Rather than
define infrared spectral types, however, Delfosse \etal\ made
use of the fact that cooling brown dwarfs slide {\em along}
an extension of the main sequence in the H-R diagram (see e.g.
D'Antona \& Mazzitelli 1985;  Burrows, Hubbard \& Lunine 1989; 
Burrows \etal\ 1997)  to classify their spectra directly
onto an M$_K$ scale. 

The results of this are shown in Table \ref{table2}
for the Mini-survey dwarfs for which optical spectroscopy has been
obtained, alongside the optical spectral types derived in Sect.\ref{spectra}.
First we can see that the three latest/faintest objects are classified as such
by both schemes, and second that the ordering of the spectra produced
by both schemes is very similar. This similarity can also be seen in a comparison of
the PC3 optical typing index of Mart\'\i n \etal\ (1996), with the
infrared spectral types.

We conclude that the procedures adopted
by Delfosse \etal\ produce results which are believable and consistent
with those produced by optical spectroscopy. At present optical spectroscopy
for the complete Mini-survey sample is not available. However, the infrared
spectra of Delfosse \etal\  would suggest that there are several
more dwarfs as late, or later, than the latest M-dwarf presented here 
(DENIS-P\,J0021). In particular, the infrared spectrum of DENIS-P\,J1228-2415
suggests it may be an M-dwarf similar to the latest M-dwarf known 2MASSP\,J0345
(Kirkpatrick \etal\ 1997a).

\section{Conclusion}

We have shown that optical spectroscopy confirms the luminosity
classifications derived from the infrared spectra of brown dwarf
candidates by Delfosse \etal\ (1998). In particular we confirm the
detection of at least two objects which are as late as the latest
known field M-dwarfs. The coolest objects studied in this work show
prominent features of Cs\,I, K\,I, CrH and FeH, and a sequence of
decreasing strength in TiO and VO features with decreasing
temperature. The coolest object in our sample also shows decreased
strength in CrH. The ``on-off'' behaviour of CrH, which is
driven by the equilibria of molecular condensation as effective
temperature decreases, implies it may be useful as a temperature diagnostic
which can be readily modelled. Lines of the K\,I doublet become
{\em extremely} strong ($\sim$300\,\AA\ equivalent width) in the
coolest DENIS objects (T$_eff \simlt 1800$K), and also offer the possibility of
being powerful temperature diagnostics. And lastly, the improved match
of models assuming rapid gravitational settling of dust grains, over
models with no settling of dust grains, would seem to indicate that
settling effects play a role in 2000-1000\,K photospheres.\\

It is clear that an entirely new sequence of spectral sub-classes
for the ``L''-type dwarfs will be required. Assignment of specific
sub-classes must await both improvements in atmospheric models to detail
the effects of dust condensation and settling on radiative transfer and the equation
of state, and an increase in the sample of L-dwarfs with
optical spectra. The latter will be achieved by the 2MASS and DENIS
surveys within the next two to three years, promising exciting
developments in our understanding of the behaviour and
properties of brown dwarfs.

\begin{acknowledgements}
We are grateful to Davy Kirkpatrick for pointing out the importance
of CrH in L-dwarf spectra prior to his own publication of the
identification, and to Brian Boyle for the award of Director's
Discretionary time on the AAT. X.D. acknowledges a
``Lavoisier'' grant of the French Minist\`ere des Affaires \'Etrang\`eres.
F.A acknowledges the support of NASA grants 110-96LTSA and NAG-3435 to 
Wichita State University.
\end{acknowledgements}


\end{document}